\documentclass[12pt]{article}
\usepackage{amsmath,epsf,amssymb,latexsym,cite}
\newif\iffigs\figstrue

\def\pplogo{\vbox{\kern-\headheight\kern -29pt
\halign{##&##\hfil\cr&{
\ppnumber}\cr\rule{0pt}{2.5ex}&\ppdate\cr}
}}
\makeatletter
\def\ps@firstpage{\ps@empty \def\@oddhead{\hss\pplogo}%
  \let\@evenhead\@oddhead 
}
\def\maketitle{\par
 \begingroup
 \def\thefootnote{\fnsymbol{footnote}}
 \def\@makefnmark{\hbox{$^{\@thefnmark}$\hss}}
 \if@twocolumn
 \twocolumn[\@maketitle]
 \else \newpage
 \global\@topnum\z@ \@maketitle \fi\thispagestyle{firstpage}\@thanks
 \endgroup
 \setcounter{footnote}{0}
 \let\maketitle\relax
 \let\@maketitle\relax
 \gdef\@thanks{}\gdef\@author{}\gdef\@title{}\let\thanks\relax}
\makeatother

\def\C{{\mathbb C}}

\def\Z{{\mathbb Z}}

\def\Hom{\operatorname{Hom}}

\def\GU{\operatorname{U{}}}

\def\labto#1{\mathrel{\mathop\to^{#1}}}

\begin{document}
\setcounter{page}0
\def\ppnumber{\vbox{\baselineskip14pt\hbox{DUKE-CGTP-00-17}
\hbox{hep-th/0009045}}}
\def\ppdate{September 2000} \date{}

\title{\LARGE A Note on the Equivalence of Vafa's and Douglas's
Picture of Discrete Torsion\\[10mm]}
\author{
Paul S. Aspinwall\\[10mm]
\normalsize Center for Geometry and Theoretical Physics, \\
\normalsize Box 90318, \\
\normalsize Duke University, \\
\normalsize Durham, NC 27708-0318\\[10mm]
}

{\hfuzz=10cm\maketitle}

\def\Large{\large}
\def\LARGE{\large\bf}


\begin{abstract}
For a general nonabelian group action and an arbitrary genus
worldsheet we show that Vafa's old definition of discrete torsion
coincides with Douglas's D-brane definition of discrete torsion
associated to projective representations.
\end{abstract}

\vfil\break


\section{Vafa's picture}    \label{s:Vafa}

Discrete torsion was introduced many years ago by Vafa in
\cite{Vafa:tor}. More recently Douglas \cite{D:disctor} introduced an
alternative picture in terms of D-branes. The purpose of this note is
to show the equivalence of these pictures. Since both of these papers
were very brief as regards to the general case we will review both
constructions here.

In general a worldsheet $\Sigma$ of genus $g$ and no boundary will be
associated to a given phase dictated by the $B$-field and the homology
class of the worldsheet in the target space. Vafa wanted to generalize
this notion of a $B$-field to the case of
orbifolds with fixed points. 

In order to localize the picture let us consider a discrete group $\Gamma$
acting on $\C^n$ where the origin is fixed by all of $\Gamma$. The
worldsheet $\Sigma$ on this orbifold may be pictured as a disk in the
covering $\C^n$ where the edges of the disk are identified by elements
of the action of $\Gamma$. That is, the usual homology 1-cycles of the
worldsheet $\Sigma$ are twisted by elements of $\Gamma$.

The twisting of these cycles in $\Sigma$ may be viewed as a
$\Gamma$-bundle on $\Sigma$. The holonomy of this bundle is given
precisely by the twist associated to a given loop. Since $\Gamma$ is a
discrete group, this bundle is flat. Such a bundle is classified by
the homotopy class of a mapping $\phi:\Sigma\to B\Gamma$, where
$B\Gamma$ is the classifying space for the group $\Gamma$. That is,
$\pi_1(B\Gamma)=\Gamma$ and $\pi_n(B\Gamma)=0$ for $n>1$.

In this way, $B\Gamma$ in the case of an orbifold plays a rather
similar role to the target space of the usual string theory. Vafa's
idea was to define the discrete torsion version of the $B$-field as 
$H^2(B\Gamma,\GU(1))$ because of this analogy. The connection between
the $B$-field and group cohomology has been clarified recently by the
work of Sharpe \cite{Sh:gerbDT}.

Now group (co)homology, which is defined as the (co)homology of
$B\Gamma$, can be described algebraically directly in terms of
$\Gamma$ as we now show. In order to complete the description of
discrete torsion we need to describe the above picture in terms of
this more intrinsic definition.

Let $\Z\Gamma$ be the {\em group ring\/} of $\Gamma$. That is, any
element of $\Z\Gamma$ may be written uniquely as
$\sum_{g\in\Gamma}a_gg$ for $a_g\in\Z$.
Now let
\begin{equation}
  \ldots\to F_n\to\ldots\to F_1\to F_0\to\Z\to0,
	\label{eq:ZGfree}
\end{equation}
be a free resolution of $\Z$ as a $\Z\Gamma$-module. This is
an exact sequence of $\Z\Gamma$-modules where $F_n$ is a free
$\Z\Gamma$-module for any $n$. The $\Gamma$-action on $\Z$ is taken to
be trivial.

Now define $(F_n)_\Gamma$ as the $\Gamma$-coinvariant projection of
$F_n$. That is, we divide $F_n$ by the equivalence $g\cong1$ for any
$g\in\Gamma$. Since 
$F_n$ is a free $\Z\Gamma$-module, $(F_n)_\Gamma$ will be a free
$\Z$-module, i.e., a free abelian group. The homology of the induced
complex
\begin{equation}
  \ldots\to (F_n)_\Gamma\to\ldots\to (F_1)_\Gamma\to (F_0)_\Gamma\to0
	\label{eq:free}
\end{equation}
is then equal to $H_n(\Gamma)$, the homology of $\Gamma$. See section
II.4 of \cite{Brown:}, for example, for a proof that this equals
$H_n(B\Gamma)$. 

One way to explicitly compute $H_n(\Gamma)$ is via the ``bar
resolution'' as follows.\footnote{In practice this method is usually
very inefficient!} Let $F_n$ be generated by $(n+1)$-tuples of the
form $(g_0,g_1,\ldots,g_n)$ where the $\Gamma$-action is defined as 
\begin{equation}
  g:(g_0,g_1,\ldots,g_n)\mapsto(gg_0,gg_1,\ldots,gg_n).
\end{equation}
The boundary map $\partial:F_n\to F_{n-1}$ is defined as
\begin{equation}
  \partial:(g_0,g_1,\ldots,g_n)\mapsto\sum_{i=0}^n(-1)^i
	(g_0,g_1,\ldots,\widehat{g_i},\ldots,g_n),
\end{equation}
where the hat indicates omission as usual. Because of the
$\Gamma$-action, we may use $(1,g_1,\ldots,g_n)$ as a basis for 
$F_n$ or $(F_n)_\Gamma$. The ``bar'' notation is to write
\begin{equation}
  [g_1|g_2|\ldots|g_n] = (1,g_1,g_1g_2,\ldots,g_1g_2\cdots g_n).
\end{equation}
We will generally consider the homology of $\Gamma$ in
terms of the generators $[g_1|g_2|\ldots|g_n]$. One may show that a
generator may be considered trivial if any of the entries
$g_1,g_2,\ldots$ are equal to 1. 
It is also useful to note the explicit form of the boundary map in
(\ref{eq:free}) of a 3-chain in the bar notation:
\begin{equation}
\partial[a|b|c] = [b|c]-[ab|c]+[a|bc]-[a|b].   \label{eq:3c}
\end{equation}

So how is the homology class of $\Sigma$ in $B\Gamma$ described in
terms of these bar chains? The free resolution (\ref{eq:ZGfree}) is
actually an augmented simplicial chain complex, $\Delta$, as
follows. Let the elements of $\Gamma$ be viewed as the vertices of
$\Delta$. Now let $(g_0,g_1,\ldots,g_n)$ be the $n$-dimensional
simplex in $\Delta$ with the corresponding vertices. Thus we have
exactly one simplex in $\Delta$ for every possible sequence of group
elements. The group $\Gamma$ acts on $\Delta$ to give the complex
(\ref{eq:free}) as a quotient. As far as homology is concerned, the
abstract simplicial complex $\Delta/\Gamma$ is a perfectly good
representative for $B\Gamma$.

\iffigs
\begin{figure}
  \centerline{\epsfxsize=5cm\epsfbox{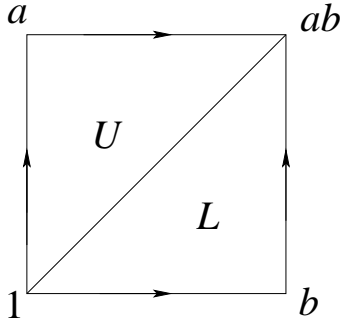}}
  \caption{A simplicial decomposition of a torus.}
  \label{fig:torus}
\end{figure}
\fi

We may now take a simplicial decomposition of $\Sigma$ and explicitly
map it into this simplicial picture of $B\Gamma$. This will relate the
homology of $\Sigma$, and in particular its fundamental class, to the
homology of $\Gamma$. As a simple example let us consider a torus as
shown in figure~\ref{fig:torus}. This torus consists of one cycle
twisted by the action of $a\in\Gamma$ and another cycle twisted by the
action of $b\in\Gamma$. The topology of the torus dictates that
$ab=ba$. The figure shows the labels of the vertices in terms of the
group action and it also shows a simple simplicial decomposition. We are free to label one vertex as ``1''. In terms of
the image of this torus in our peculiar simplicial model of $B\Gamma$
we see that $U$ corresponds to a simplex $(1,a,ab)$ and $L$
corresponds to a simplex $(1,b,ab)$. As usual in simplicial homology
one needs to be careful about relative signs. The fundamental class of
$\Sigma$ is given by $U-L$ so that the diagonal line in
figure~\ref{fig:torus} cancels for the boundary of $U-L$. This gives
\begin{equation}
\begin{split}
  [\Sigma] &= (1,a,ab)-(1,b,ab)\\
           &= [a|b]-[b|a],   \label{eq:g1}
\end{split}
\end{equation}
in terms of the bar chains. 

Applying $\Hom(-,\GU(1))$ to this bar resolution 
we may go over to the group cohomology picture. A group $n$-cochain
can now be viewed as a map from $(\Gamma)^n$ to $\GU(1)$. For example, if
$\alpha\in H^2(\Gamma,\GU(1))$ then 
$\alpha(a,b)\in\GU(1)$ for $a,b\in\Gamma$. From (\ref{eq:3c}) we see
that a 2-cochain is closed if
\begin{equation}
\begin{split}
  (\delta\alpha)(a,b,c) &= \frac{\alpha(a,bc)\alpha(b,c)}
	{\alpha(a,b)\alpha(ab,c)}\\
	&= 1.
\end{split}  \label{eq:coc2}
\end{equation}

If discrete torsion is written in terms of the $\alpha$ cocycles then 
(\ref{eq:g1}) is translated in Vafa's language into the statement that
the phase associated to a genus one Riemann surface is given by
\begin{equation}
  \xi_1 = \frac{\alpha(a,b)}{\alpha(b,a)}.
\end{equation}

\iffigs
\begin{figure}
  \centerline{\epsfxsize=5cm\epsfbox{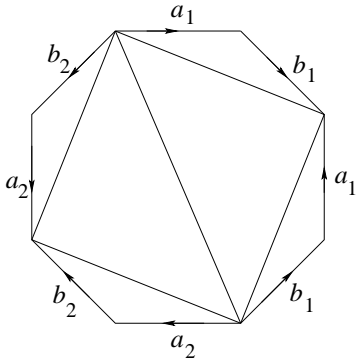}}
  \caption{A simplicial decomposition of a genus 2 surface.}
  \label{fig:g2}
\end{figure}
\fi

This is the result quoted in
\cite{Vafa:tor}. This method may be applied
equally well to higher genus worldsheets. The genus two case is shown
in figure~\ref{fig:g2} together with a specific simplicial
decomposition. It follows that
\begin{equation}
  [\Sigma] = [a_1|b_1]-[\gamma|b_1a_1]-[b_1|a_1]+[a_2|b_2]
               +[\gamma|a_2b_2]-[b_2|a_2],
\end{equation}
where $\gamma=a_1b_1a_1^{-1}b_1^{-1}$, which by the topology of a
genus two surface must equal $b_2a_2b_2^{-1}a_2^{-1}$. Note that if
$\gamma=1$, which would happen if $\Gamma$ were abelian for example,
then the genus two homology class simply breaks into
$([a_1|b_1]-[b_1|a_1])+([a_2|b_2]-[b_2|a_2])$ which is the sum of two
genus one classes. This does {\em not\/} happen in general if
$\gamma\neq1$. One may view this statement in terms of a genus two
surface degenerating into two genus one surfaces touching at a
point. Such a degeneration is topologically obstructed if $\gamma\neq1$.

There are many ways of performing a simplicial decomposition of
$\Sigma$ in general. One may show that changing the decomposition will
simply change $[\Sigma]$ in the computations above by the boundary of
a group 3-chain and hence has no effect.


\section{Douglas's picture}    \label{s:Doug}

In \cite{D:disctor} Douglas introduced another definition of discrete
torsion associated to projective representations which we now review. See
\cite{Gomis:dt,AP:mck} for further discussion of this construction.

Consider the following central extension of $\Gamma$:
\begin{equation}
\renewcommand{\arraystretch}{0.1}
  1 \to \GU(1) \labto{i} \hat\Gamma\begin{array}{c}{\scriptstyle s}\\ 
  \curvearrowleft\\{\scriptstyle j}\\\to\\ 
  \vphantom{l}\\ \vphantom{m}\end{array}\Gamma \to 1,
\end{equation}
where $s$ is a set-theoretic map such that $js$ is the identity on
$\Gamma$. Such extensions are classified by $H^2(\Gamma,\GU(1))$.
The map $s$ defines a projective representation of $\Gamma$. Given
$\alpha\in H^2(\Gamma,\GU(1))$ written in terms of the bar resolution
of the previous section one may show that \cite{Karpil:proj}
\begin{equation}
  s(a)s(b) = \alpha(a,b)s(ab).  \label{eq:proj}
\end{equation}

Douglas considered $s$ as a lift of an orbifold action to the
Chan-Paton factors on the end of an open string. He then considered a
Riemann surface with a disk removed. Such a surface can represent
either a multi-loop open string diagram or a multi-loop tadpole-like
diagram for a closed string. 

\iffigs
\begin{figure}
  \centerline{\epsfxsize=5cm\epsfbox{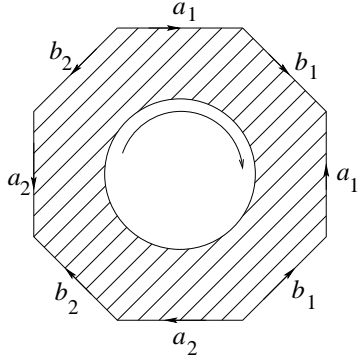}}
  \caption{A genus 2 surface with a hole cut out.}
  \label{fig:2a}
\end{figure}
\fi

Consider the genus two case shown in figure~\ref{fig:2a}. When
computing the amplitude of this diagram in the open string context one
must include a trace of the group actions on the Chan-Paton
elements along the boundary. As is clear from the figure, 
this boundary may be contracted to the outer polygon which is a
sequence of 1-cycles which are twisted by elements of
$\Gamma$. Let $\xi_2$ be this weighting of the amplitude. We then have
\begin{equation}
  \xi_2 = s(a_1)s(b_1)s(a_1)^{-1}s(b_1)^{-1}s(a_2)s(b_2)
	s(a_2)^{-1}s(b_2)^{-1},
\end{equation}
where $a_i,b_i\in\Gamma$ represent the associated twists. For a
general genus $g$ surface we clearly have
\begin{equation}
  \xi_g = \prod_{i=1}^g s(a_i)s(b_i)s(a_i)^{-1}s(b_i)^{-1}.
\end{equation}
Let $\gamma_i=a_ib_ia_i^{-1}b_i^{-1}$. Then the topology of the
surface dictates that $\prod_{i=1}^g\gamma_i=1$.

\iffigs
\begin{figure}
  \centerline{\epsfxsize=5cm\epsfbox{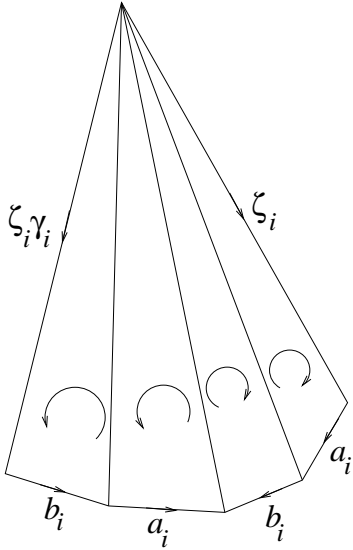}}
  \caption{Part of the simplicial decomposition of the arbitrary genus case.}
  \label{fig:gen}
\end{figure}
\fi

Note that $s(1)=1$ and thus $s(x)s(x^{-1})=\alpha(x,x^{-1})$. It
follows that $\alpha(x,x^{-1})=\alpha(x^{-1},x)$. Repeated use of
(\ref{eq:proj}) and (\ref{eq:coc2}) then gives\footnote{We assume
$g>1$. The case $g=1$ is left as an easy exercise for the reader.}
\begin{equation}
\begin{split}
\xi_g &= \prod_{i=1}^g \frac{s(a_i)s(b_i)s(a_i^{-1})s(b_i^{-1})}
	{\alpha(a_i,a_i^{-1})\alpha(b_i,b_i^{-1})}\\
      &=\frac{\alpha(a_1,b_1)\alpha(a_1b_1,a_1^{-1})
		\alpha(a_1b_1a_1^{-1},b_1^{-1})\ldots
		\alpha((\prod_{i=1}^{g-1}\gamma_i)a_gb_ga_g^{-1},b_g^{-1})}
	{\prod_{i=1}^g\alpha(a_i,a_i^{-1})\alpha(b_i,b_i^{-1})}\\
      &=\frac{\alpha(a_1,b_1)\alpha(\gamma_1b_1a_1,a_1^{-1})
	\alpha(\gamma_1b_1,b_1^{-1})\ldots
		\alpha(b_g,b_g^{-1})}
	{\prod_{i=1}^g\alpha(a_i,a_i^{-1})\alpha(b_i,b_i^{-1})}\\
      &=\frac{\alpha(a_1,b_1)}{\alpha(\gamma_1b_1,a_1)\alpha(\gamma_1,b_1)}
	\cdot
	\prod_{i=2}^{g-1}\frac{\alpha(\zeta_i,a_i)
		\alpha(\zeta_ia_i,b_i)}
		{\alpha(\zeta_i\gamma_ib_i,a_i)
			\alpha(\zeta_i\gamma_i,b_i)}\cdot
	\frac{\alpha(\zeta_g,a_g)\alpha(\zeta_ga_g,b_g)}{\alpha(b_g,a_g)},
\end{split}   \label{eq:big}
\end{equation}
where $\zeta_i=\gamma_1\gamma_2\cdots\gamma_{i-1}$. 

This formula was also derived in \cite{Ban:dt}.
Now it is not hard to see that the final form of (\ref{eq:big})
corresponds to a simplicial decomposition of a genus $g$ surface into 
$4g-2$ triangles in the language of section~\ref{s:Vafa}. Each
$\alpha$ factor represents one triangle 
oriented in just the right way to build up the complete surface. We
show the four triangles for a generic pair $a_i,b_i$ in
figure~\ref{fig:gen}.

This factor $\xi_g$ is exactly the same factor as we obtain by
Vafa's method of the previous section applied to this simplicial
decomposition. Thus we see that Vafa's and 
Douglas's definition of discrete torsion agree in general.


\section*{Acknowledgements}

It is a pleasure to thank S.~Katz, R.~Plesser and E.~Sharpe for useful
conversations. 
The author is supported in part by NSF grant DMS-0074072 and a research
fellowship from the Alfred P.~Sloan Foundation. 


\end{document}